\documentclass[aps,prl,twocolumn,superscriptaddress,showpacs]{revtex4}
\usepackage{epsfig}
\usepackage{amsmath}
\usepackage{graphicx}

\begin{document}

\title{Domain enhanced interlayer coupling in ferroelectric/paraelectric
superlattices}
\author{V. A. Stephanovich}
\affiliation{University of Opole, Institute of Mathematics and Informatics,Opole,45-052,
Poland}
\author{I. A. Luk'yanchuk}
\affiliation{University of Picardie Jules Verne, Laboratory of Condensed Matter Physics,
Amiens, 80039, France}
\affiliation{L. D. Landau Institute for Theoretical Physics, RAS, 117940 GSP-1, Moscow,
Russia}
\author{M. G. Karkut}
\affiliation{University of Picardie Jules Verne, Laboratory of Condensed Matter Physics,
Amiens, 80039, France}
\date{\today}

\begin{abstract}
We investigate the ferroelectric phase transition and domain formation in a
periodic superlattice consisting of alternate ferroelectric (FE) and
paraelectric (PE) layers of nanometric thickness. We find that the
polarization domains formed in the different FE layers can interact with
each other via the PE layers. By coupling the electrostatic equations with
those obtained by minimizing the Ginzburg-Landau functional we calculate the
critical temperature of transition $T_c$ as a function of the FE/PE
superlattice wavelength $\Lambda$ and quantitatively explain the recent
experimental observation of a thickness dependence of the ferroelectric
transition temperature in KTaO$_{3}$/KNbO$_{3}$ strained-layer superlattices.
\end{abstract}

\pacs{\ 77.55.+f,\ 77.80.Dj,\ 77.80.Bh}
\maketitle




In the past decade refinements in deposition techniques have made it
possible to fabricate nanoscale size oxide ferroelectric superlattices with
the objective to merge and optimize the technological properties of the
constitutive materials \cite{Scott,Hong,Ishiwara}. In designing such
artificial structures an understanding of the physics of underlying
processes is essential to determine whether the resulting characteristics
are provided simply by the superposition of the bulk properties of the
constituents or whether the interface and finite-size effects play a
predominant role.

Two competing types of phenomena that arise at the ferroelectric interface
can affect the properties of the superlattices. The strain field, generated
by the mechanical mismatch between the superlattice layers, influences the
polarization orientation and generally increases the ferroelectric
transition temperature $T_{c}$ \cite{Pertsev}. In contrast, the electric
depolarization field, produced by interfacial surface charges is unfavorable
to the formation of the ferroelectric phase \cite{Batra}. In fact, in cubic
perovskite-like ferroelectrics the situation can be even more complex due
the formation of both $180^\circ$ ferroelectric \cite{bratlev} and $90^\circ$
ferroelastic \cite{Roitburd,Kwak,Pertsev} domains. Although the properties
of ferroelectric superlattices can be governed by domain structure, no
systematic study of this effect has to our knowledge been performed.

In the present paper, we address the question of ferroelectric domain
formation in a periodic superlattice structure consisting of alternate
ferroelectric (FE) and paraelectric (PE) layers of equal nanometric width $%
2a_{f}=2a_{p}$. So as to avoid the complications of the effect of $90^{\circ
}$ ferroelastic domains we assume that the ferroelectric layers have either
natural or strain-induced c-oriented uniaxial symmetry. We will show that
the domain patterns formed in the different FE layers interact with each
other across the PE layers via the spatially inhomogeneous depolarization
electric field emerging from the domains of the neighboring FE layers as
shown in Fig. 1. This proximity type effect is dependent critically on the
thickness of the PE layers. Our interest has also been motivated by a recent
experimental study of FE/PE superlattices of KTaO$_{3}$/KNbO$_{3}$ \cite%
{Specht} in which, as the superlattice wavelength $\Lambda
=2a_{f}+2a_{p}$ decreases, the ferroelectric transition temperature
$T_{c}$ first decreases and then saturates below a certain layer
thickness. Such dependence was reproduced using molecular dynamics
simulations \cite{Sepliarsky} but the role of depolarizing effects
was not elucidated. Attributing the observed behavior of critical
temperature to the domain
coupling in FE layers across the PE layers, we calculate the dependence $%
T_{c}(\Lambda )$ and show that this dependence correctly reproduces the
experimental behavior.

\begin{figure}[!bt]
\vspace{5mm} \centering
\includegraphics [width=8cm]
{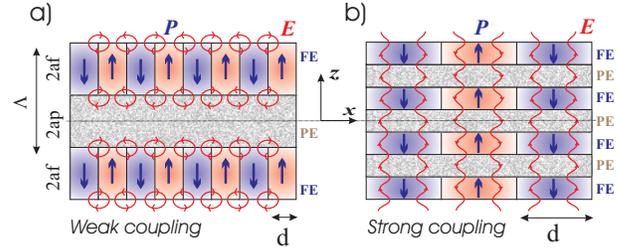}
\caption{Domains and depolarization field in weak (a) and strong (b) coupled
ferroelectric layers in FE/PE superlattice close to $T_c$}
\label{Fig1_SuperlatticeDomains}
\end{figure}

The physics underlying the domain formation and interaction is as follows.
It is well known that a ferroelectric slab or thin film of thickness $2a_{f}$
will separate into domains. This is to reduce the energy of the
depolarization field produced by the space and surface charges with charge
density $\rho (r)=\mathrm{div}\mathbf{P}$ that are provided by the
discontinuity and non-uniformity of the polarization close to the crystal
surface. In the Kittel approximation \cite{Kittel,Landau8,StrukovLevanyuk},
applicable at $T \ll T_{c}$, the polarization inside a domain is assumed to
be uniform and almost equal to its equilibrium value $P_{0}$. Thus the
depolarization field is proportional to the polarization discontinuity $%
E_{\sigma }\sim 4\pi P_{0}$ and is confined to a thin layer of penetration
length $\delta $ that is roughly proportional to the domain size $d$. The
equilibrium domain structure is the result of a balance between two
competing energies (calculated per unit surface of crystal): the
depolarization energy $F_{\sigma }\sim \delta E_{\sigma }^{2}\sim dP_{0}^{2}$
and the energy due to polarization gradients inside the domain walls $%
F_{w}\sim a_{f}\Delta P_{0}^{2}n\sim \Delta P_{0}^{2}(a_{f}/d)$ where the
length scale parameter $\Delta $ is called a "domain wall width" and $n\sim
1/d$ is the wall concentration. Minimization of $F_{\sigma }+F_{w}$ gives
the famous Kittel formula: $d\sim \sqrt{\Delta a_{f}}$ \cite{Kittel,Landau8}.

The coupling between FE layers in a FE/PE superlattice is caused by the
depolarization field emerging from the domain structure of the FE layers.
This interaction is exponentially small and the FE layers are almost
independent if the distance between them $2a_{p}$ is larger then the
penetration length $\delta $. Taking into account that $\delta $ scales as $%
a_{f}^{1/2}$, we determine that for an equally layered superlattice with $%
2a_{p}=2a_{f}=\Lambda /2$ this weak coupling regime is realized for long
wavelength $\Lambda $. In the opposite limit of short $\Lambda $, the
depolarization field penetrates the PE layers and so couples the domains in
the neighboring ferroelectric layers. In this strongly coupled regime the
domain size exceeds the superlattice wavelength and the superlattice behaves
effectively as a uniform "composite" ferroelectric.

Close to the ferroelectric transition the polarization varies gradually
inside domains, the charge and depolarization field penetrate FE layers so
that the Kittel approximation is no longer applicable. In order to determine
the domain structure parameters in FE/PE superlattices close to $T_{c}$ and
also the domain-formation-induced reduction of $T_{c}$ with respect to the
bare $T_{c0}$ of a bulk sample in which the depolarization field is screened
by the short circuited electrodes, we generalize the analogous calculations
\cite{chentar} for a thick ferroelectric plate. We start from the complete
system of electrostatic equations:
\begin{equation}
\mathrm{div}(\mathbf{E}+4\pi \mathbf{P})=0,\quad \mathrm{rot}\ \mathbf{E}=0,
\label{e1}
\end{equation}%
in which the polarization is related to the electric field by the
constitutive relation $\mathbf{P}=\mathbf{P}(\mathbf{E})$. This is to be
determined from both the nonlinear Ginzburg-Landau equation for the $z$%
-component of the spontaneous polarization in the uniaxial ferroelectric
layers $P=P_{z}^{(f)}$:
\begin{equation}
tP+P_{0}^{-2}P^{3}-\xi _{0}^{2}\,\nabla ^{2}P=\frac{\varepsilon _{\parallel }%
}{4\pi }E_{z}^{(f)},  \label{GL}
\end{equation}%
(here $t=T/T_{c0}-1$, $\nabla ^{2}=\partial _{z}^{2}+\partial _{x}^{2}$) and
by the linear relations for the two polarization components $P_{x,z}^{(p)}$
in the PE-layers as well as for the transversal component $P_{x}^{(f)}$ in
the FE-layers:
\begin{equation}
P_{x}^{(f)}=\frac{\varepsilon _{\perp }-1}{4\pi }E_{x}^{(f)},\quad
P_{x,z}^{(p)}=\frac{\varepsilon _{p}-1}{4\pi }E_{x,z}^{(p)}.  \label{polar}
\end{equation}%
The susceptibility $\varepsilon _{p}$ of the PE-layers in (\ref{polar}) is
assumed to be isotropic. The dimensionless parameter $\varepsilon
_{\parallel }\gg 1$ in (\ref{GL}) is expressed via the Curie constant as $%
\varepsilon _{\parallel }=C/T_{c0}$. The length $\xi _{0}\simeq 6\mathring{A}
$ is estimated as the domain wall half-width at low temperatures, i.e. well
below the phase transition. Hereafter we shall scale all the lengths in
units of $\xi _{0}$.

The formulation of the problem is completed by the electrostatic boundary
conditions at the PE/FE interfaces:
\begin{equation}
E_{z}^{(f)}-E_{z}^{(p)}=-4\pi (P_{z}^{(f)}-P_{z}^{(p)}),\quad
E_{x}^{(f)}=E_{x}^{(p)}  \label{BK}
\end{equation}%
and by the interface condition for the spontaneous polarization:
\begin{equation}
\partial _{z}P=\lambda P,  \label{lambd}
\end{equation}%
where $\lambda $ is the extrapolation length \cite{Scott,Binder} that
reflects the properties of the interface.

Close to the ferroelectric transition Eq.(\ref{GL}) can be linearized by
neglecting the term $P_{0}^{-2}P^{3}$. The dimensionless transition
temperature $t_{c}$, where the instability to the formation of a multidomain
structure first appears, can be found as the highest eigenvalue of the
linearized system of equations (\ref{e1}), (\ref{GL}) and (\ref{polar}) with
the appropriate boundary conditions (\ref{BK}) and (\ref{lambd}).

In terms of the electrostatic potential $\varphi ^{(f,p)}$: $%
E_{z,x}^{(f,p)}=-\partial _{z,x}\varphi ^{(f,p)}$, the full set of
linearized equations takes the form: \newline
\null\hspace{0.3cm} \textit{\ PE-layers: }
\begin{equation}
(\partial _{z}^{2}+\partial _{x}^{2})\varphi ^{(p)}=0.  \label{fip}
\end{equation}%
\null\hspace{0.3cm} \textit{FE-layers: }
\begin{gather}
4\pi \varepsilon _{\parallel }^{-1}(t-\,\nabla ^{2})P=-\partial _{z}\varphi
^{(f)},  \label{fif} \\
(\partial _{z}^{2}+\varepsilon _{\perp }\partial _{x}^{2})\varphi
^{(f)}=4\pi \partial _{z}P.  \label{fiff}
\end{gather}%
\null\hspace{0.3cm} \textit{PE/FE interface: }
\begin{eqnarray}
&&\partial _{z}\varphi ^{(f)}-\varepsilon _{p}\partial _{z}\varphi
^{(p)}=4\pi P,  \label{bcfi1} \\
&&\varphi ^{(f)}=\varphi ^{(p)},\quad \partial _{z}P=\lambda P.
\label{bcfi2}
\end{eqnarray}%
The eigenfunctions of the system of elliptic equations (\ref{fip}), (\ref%
{fif}) and (\ref{fiff}) are the linear superposition of the harmonic
functions which we write as:\newline
\textit{For the upper (lower) FE-layer} (see Fig. 1)
\begin{gather}
P =\left[ P_{1}\frac{\cosh k_{1}\left( z\mp {\frac{\Lambda }{2}}\right) }{%
\cosh k_{1}a_{f}}+P_{2}\frac{\cosh k_{2}\left( z\mp {\frac{\Lambda }{2}}
\right) }{\cosh k_{2}a_{f}}\right] \cos qx,  \label{solf} \\
\varphi ^{(f)} =\left[ \varphi _{1}\frac{\sinh k_{1}\left( z\mp {\frac{%
\Lambda }{2}}\right) }{ \sinh k_{1}a_{f}}+\varphi _{2}\frac{\sinh
k_{2}\left( z\mp {\frac{\Lambda }{2}}\right) }{\sinh k_{2}a_{f}}\right] \cos
qx.  \notag
\end{gather}%
\textit{For the central PE-layer}
\begin{equation}
\varphi ^{(p)}=-(\varphi _{1}+\varphi _{2})\frac{\sinh qz}{\sinh qa_{p}}\cos
qx.  \label{solp}
\end{equation}%
In (\ref{solp}) the boundary conditions (\ref{bcfi2}) were assumed. The
periodicity of the oscillating factor $\cos qx$ reflects the formation of a
regular domain structure in the $x$- direction with domain size $d = \pi /q$%
. The solutions (\ref{solf}) and (\ref{solp}) can be periodically continued
in the $z$-direction so as to follow the regular FE/PE superlattice
structure.

The parameters $P_{1,2},$ $\varphi _{1,2}$ and $k_{1,2}$ are found by
substitution of (\ref{solf}) and (\ref{solp}) back into (\ref{fip}),(\ref%
{fif}),(\ref{fiff}),(\ref{bcfi1}) and (\ref{bcfi2}). This permits us to find
two characteristic equations defining the eigentemperature $t=t(q)$.

Substitution of solutions (\ref{solf}) into Eqs. (\ref{fif}) and (\ref{fiff}%
) produces a homogeneous system of linear equations for $\varphi _{1,2}$ and
$P_{1,2}$:
\begin{eqnarray}
&&4\pi \varepsilon _{\parallel }^{-1}\left( t-k_{1,2}^{2}+q^{2}\right)
P_{1,2}+k_{1,2}\coth (k_{1,2}a_{f})~\varphi _{1,2}=0,  \notag \\
&&4\pi k_{1,2}~P_{1,2}-\left( k_{1,2}^{2}-\varepsilon _{\perp }q^{2}\right)
\coth (k_{1,2}a_{f})~\varphi _{1,2}=0,  \label{uv1b}
\end{eqnarray}
that are compatible if the characteristic equation
\begin{equation}
\left( k_{1,2}^{2}-\varepsilon _{\perp }q^{2}\right) \left(
t-k_{1,2}^{2}+q^{2}\right) +\varepsilon _{\parallel }k_{1,2}^{2}=0
\label{xar1}
\end{equation}
is satisfied.

\begin{figure}[!bt]
\vspace{5mm} \centering
\includegraphics [width=6.5cm]{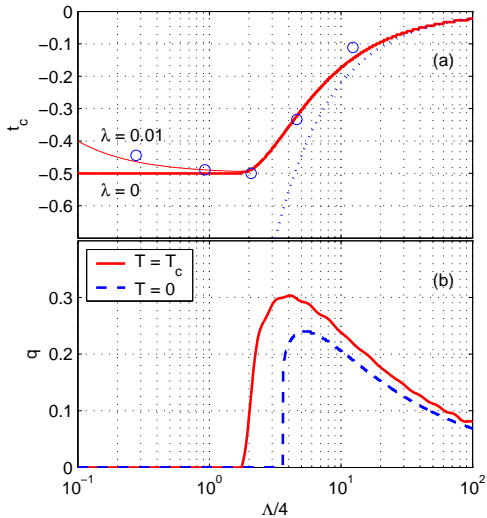}
\caption{Critical temperature $t_c=T_c/T_{c0}-1$ (a) and domain structure
wave vector $q=\protect\pi/d$ at $T=0$ and $T=T_c$ (b) as a function of
superlattice wavelength $\Lambda=2a_f+2a_p$. Dotted line shows the asymptote
(\protect\ref{syma}). Circles show the experimental data for KTaO$_{3}$/KNbO$%
_{3}$ superlattice {{\protect\cite{Specht}}}. The best fit parameters $%
\protect\varepsilon _\parallel =400$, $\protect\varepsilon _\perp =500$, $%
\protect\varepsilon _p =800$, $\protect\lambda=0$ and $\protect\lambda=0.01$
have been used. The length is scaled in units of $\protect\xi _{0}\simeq 6%
\mathring{A} $}
\label{Fig2_SuperlatticeDomains}
\end{figure}

Now, to obtain the $t(q)$ dependence we need one more equation relating $t$,
$k$, and $q$. This equation follows from the boundary conditions. Condition (%
\ref{bcfi1}) gives the following relation between $\varphi _{1,2}$ and $%
P_{1,2}$:
\begin{equation}
Q_{1}\coth k_{1}a_{f}~\varphi _{1}+Q_{2}\coth k_{2}a_{f}~\varphi _{2}=4\pi
\left( P_{1}+P_{2}\right) ,  \label{u10a}
\end{equation}
with $Q_{1,2}=k_{1,2}+\varepsilon _{p}q\coth qa_{p}\tanh k_{1,2}a_{f}$,
while condition (\ref{lambd}) gives:
\begin{equation}
P_{1}(k_{1}\tanh k_{1}a_{f}-\lambda )+P_{2}(k_{2}\tanh k_{2}a_{f}
-\lambda)=0.  \label{u10b}
\end{equation}
Using (\ref{uv1b}) we can express $\varphi _{1,2}$ as a function of $P_{1,2}$%
:
\begin{equation}
\varphi _{1,2}=P_{1,2}\frac{4\pi k_{1,2}}{k_{1,2}^{2}-\varepsilon _{\perp
}q^{2}}\tanh k_{1,2}a_{f}.  \label{u11}
\end{equation}
Substitution of (\ref{u11}) into (\ref{u10a}) yields:
\begin{equation}
P_{1}\left[ \frac{k_{1}Q_{1}}{k_{1}^{2}-\varepsilon _{\perp }q^{2}}-1\right]
+P_{2}\left[ \frac{k_{2}Q_{2}}{k_{2}^{2}-\varepsilon _{\perp }q^{2}}-1\right]
=0.  \label{ui}
\end{equation}
The compatibility criterion of equations (\ref{ui}) and (\ref{u10b}) gives
the second characteristic equation
\begin{widetext}
\begin{equation}
\frac{k_1^2-\varepsilon _{\perp }q^2}{k_2^2-\varepsilon _{\perp }q^2}\ \frac{%
\varepsilon _{\perp }q\tanh qa_p+\varepsilon _pk_2\tanh
k_2a_f}{\varepsilon _{\perp }q\tanh qa_p+\varepsilon _pk_1\tanh
k_1a_f}=\frac{k_2\tanh k_2a_f-\lambda }{k_1\tanh k_1a_f-\lambda }.
\label{xar2}
\end{equation}
\end{widetext}
The eigentemperature $t(q)$, transition temperature $t_{c}=\max_{q}t(q)$ and
the corresponding domain structure wave vector $q_{c}$ are found from
equations (\ref{xar1}) and (\ref{xar2}) after eliminating the variables $%
k_{12}$. The numerically obtained results for $t_{c}$ and $q_{c}$ as a
function of superlattice wavelength $\Lambda$ are plotted in Fig. 2 together
with the experimental results of Ref. \cite{Specht} on $t_{c}(\Lambda)$ in
KTaO$_{3}$/KNbO$_{3}$ superlattice. The parameters we used to achieve this
remarkably good fit were: $\varepsilon _{\parallel }=400$, $\varepsilon
_{\perp }=500$, $\varepsilon _{p}=800$, $\lambda =0$ and $\lambda =0.01$.

Two regimes that correspond to the above described weak and strong coupling
limits are clearly seen in Fig.~2. With decreasing $\Lambda$ the domain size
decreases, goes through a minimum and then diverges.

In the \textit{weak coupling regime} ($\Lambda >20)$ the domain size is
smaller then the superlattice wavelength and the depolarization field is
confined essentially at the FE/PE interfaces. Considering each FE layer as
an independent thick film embedded in the PE media  and using the
substitution $p_{2}=ik_{2}$ (so that $k_{2}\tanh k_{2}a_{f}=-p_{2}\tan
p_{2}a_{f}$) we can simplify Eq. (\ref{xar2}) to the form:
\begin{equation}
\varepsilon _{\perp }q\tanh qa_{p}-\varepsilon _{p}p_{2}\tan p_{2}a_{f}
\approx 0  \label{Chen}
\end{equation}%
that was first worked out in \cite{chentar}. Using Eqs. (\ref{xar1}) and (%
\ref{Chen}) the behavior of $p_{2}$, $q_{c}^2$ and $t_{c}$ reduces to:
\begin{equation}
p_{2}\approx \frac{\pi }{2a_{f}},\quad q_{c}^{2}\approx \sqrt{\frac{%
\varepsilon _{\parallel }}{\varepsilon _{\perp }}}\frac{\pi }{2a_{f}},\quad
t_{c}\approx -\sqrt{\frac{\varepsilon _{\parallel }}{\varepsilon _{\perp }}}%
\frac{\pi }{a_{f}}.  \label{syma}
\end{equation}%
The critical temperature $t_c$ is inversely proportional to $a_{f}$, exactly
as was found numerically in \cite{Wang}.

In the \textit{strong coupling regime}, below a certain critical thickness ($%
\Lambda <5)$ the domain structure abruptly disappears. The transition
temperature of single-domain FE/PE superlattice is calculated from (\ref%
{xar1}) and (\ref{xar2}) by assuming $q=0$:
\begin{equation}
t_{c}\approx -\frac{\varepsilon _{_{\parallel }}}{\varepsilon _{p}}+\frac{%
\lambda }{a_{f}},  \label{as2}
\end{equation}%
or, equivalently, from the Landau energy of a periodic FE/PE structure with
no depolarizing surface charges at the FE/PE interfaces. The FE/PE
superlattice behaves as a uniform composite ferroelectric with a critical
temperature (\ref{as2}), greater than that of the individual FE layers (\ref%
{syma}). The positive surface $\lambda$-term tends to increase $t_c$ and
this is possibly the reason for the slight increase of the transition
temperature in the KTaO$_{3}$/KNbO$_{3}$ superlattice at very small $\Lambda$
\cite{Specht}, as shown in Fig. 2. Note that although the ferroelectric
domains can exist in the strong coupling regime, their size is larger then
the superlattice wavelength $\Lambda$ and is defined by the global
depolarization field of the sample.

To estimate the temperature evolution of the domain structure we express its
energy at $T=0$ (i.e. at $t=-1$) in the Kittel approximation, assuming an
abrupt structure of the domain wall and a flat polarization profile inside
the domains. The calculations are practically the same as those performed
for the domain structure of the uniaxial ferroelectric film surrounded by
paraelectric passive layers and embedded in a short circuited capacitor \cite%
{bratlev}. The free energy is the sum of the domain walls energy and the
electrostatic contributions:
\begin{equation}
\pi F/P_{0}^{2}=2a_{f}\Delta q+\frac{32}{q}\sum_{n=1,3,...}\frac{1}{n^{3}}%
\frac{1}{g_{n}(q)},  \label{Levan}
\end{equation}%
where
\begin{equation}
g_{n}(q)=\varepsilon _{p}\coth nqa_{p}+\sqrt{\varepsilon _{\parallel
}\varepsilon _{\perp }}\coth \left( \frac{\varepsilon _{\perp }}{\varepsilon
_{\parallel }}\right) ^{1/2}nqa_{f}  \notag
\end{equation}%
and the \textquotedblright domain wall width\textquotedblright\ $\Delta $
can be found by integration of the Landau energy of the wall as $\Delta =%
\frac{4\pi }{\varepsilon _{\parallel }}\frac{1}{4}\int_{-\infty }^{\infty
}(\tanh ^{4}(x/\sqrt{2})-1)dx\approx 8\pi \sqrt{2}/3\varepsilon _{\parallel
} $. This takes into account the fact that $P(x)=P_{0}\tanh (x/\sqrt{2})$ is
the exact single-wall solution of Eq. (\ref{GL}) at $t=-1$. The result of
the numerical minimization of (\ref{Levan}) is given by the dashed line in
Fig. 2b. When $q_{c}a_{p}$, $(\varepsilon _{\perp }/\varepsilon _{\parallel
})^{1/2}q_{c}a_{f}\gg 1$ it is approximated by the generalized Kittel
formula:
\begin{equation}
q_{c}^{2}\approx \frac{\varepsilon _{\parallel }}{\varepsilon _{p}+\sqrt{%
\varepsilon _{\parallel }\varepsilon _{\perp }}}\frac{21\zeta (3)}{2\pi
\sqrt{2}}\frac{1}{2a_{f}}\qquad \frac{21\zeta (3)}{2\pi \sqrt{2}}\approx 2.8.
\label{LK}
\end{equation}

Since, as shown in Fig. 2, the plots $q(\Lambda)$ at $T=0$ and at $T=T_c$
practically coincide, we conclude that the temperature dependence of the
domain structure wave vector is very weak. The complete calculation of the
domain structure evolution and of its dielectric constant over the entire
temperature region will be published elsewhere.

To conclude, we have demonstrated that uniaxial ferroelectric domains can
substantially influence the properties of the FE/PE superlattices. Depending
on the wavelength $\Lambda$, the superlattice can be in different domain
states. For large $\Lambda$ (typically $>5-15nm$) each FE layer has an
independent domain structure. At smaller $\Lambda$ the domains in
neighboring FE layers interact through the PE layers via the emerging
depolarization field and this results in a dramatic increase of the domain
width. In this regime the superlattice structure behaves as an effective
composite uniform ferroelectric where the large-scale domains penetrate
throughout the entire sample and are governed by the global depolarization
field.

We have calculated the ferroelectric transition temperature as function of $%
\Lambda$ and have explained the recently observed small-$\Lambda$ saturation
of $T_c$ in KTaO$_{3}$/KNbO$_{3}$ superlattice by crossover to the regime of
strongly coupled FE layers. We have also shown that the critical
superlattice crossover wavelength is nearly temperature independent.

We gratefully acknowledge the Region of Picardy and the European
Social Fund for financial support.


\end{document}